\documentclass[%
 reprint,
 amsmath,amssymb,
 aps,
 prl,
 superscriptaddress
]{revtex4-2}
\pdfoutput=1

\usepackage{aas_macros}
\usepackage{graphicx}
\graphicspath{{Figures/}}
\usepackage{dcolumn}
\usepackage{slashed}
\usepackage{bm} 
\usepackage{bbold}
\usepackage{units}  
\usepackage{todonotes}
\usepackage[colorlinks=true,citecolor=blue,urlcolor=blue,linkcolor=blue]{hyperref}

\newcommand{\MSUN}{{\rm M}_{\odot}}

\begin{document}

\title{Ultra-high energy cosmic rays deflection by \\ the Intergalactic Magnetic Field}

\author{Andrés Arámburo-García}
 \email{aramburo@strw.leidenuniv.nl}
\affiliation{%
 Institute Lorentz, Leiden University, Niels Bohrweg 2, Leiden, NL-2333 CA, the Netherlands
}

\author{Kyrylo Bondarenko}
 \email{kyrylo.bondarenko@cern.ch}
\affiliation{Theoretical Physics Department, CERN, Geneva 23, CH-1211, Switzerland}
\affiliation{L’Ecole polytechnique fédérale de Lausanne, 1015 Lausanne, Switzerland}

\author{Alexey Boyarsky}
 \email{boyarsky@lorentz.leidenuniv.nl}
\affiliation{%
 Institute Lorentz, Leiden University, Niels Bohrweg 2, Leiden, NL-2333 CA, the Netherlands
}

\author{Dylan Nelson}
 \email{dnelson@uni-heidelberg.de}
\affiliation{%
Universität Heidelberg, Zentrum für Astronomie, Institut für theoretische Astrophysik, Albert-Ueberle-Str. 2, 69120 Heidelberg, Germany
 }

\author{Annalisa Pillepich}
\email{pillepich@mpia-hd.mpg.de}
\affiliation{%
Max-Planck-Institut für Astronomie, Königstuhl 17, 69117 Heidelberg, Germany
 }

\author{Anastasia Sokolenko}
 \email{anastasia.sokolenko@oeaw.ac.at}
\affiliation{Institute of High Energy Physics, Austrian Academy of Sciences, Nikolsdorfergasse 18, 1050 Vienna, Austria}

\date{\today}

\begin{abstract}
The origin and composition of ultra-high energy cosmic rays (UHECRs) remain a mystery. The common lore is that UHECRs are deflected from their primary directions by the Galactic and extragalactic magnetic fields. Here we describe an extragalactic contribution to the deflection of UHECRs that does not depend on the strength and orientation of the initial seed field. Using the IllustrisTNG simulations, we show that outflow-driven magnetic bubbles created by feedback processes during galaxy formation deflect approximately half of all $10^{20}$~eV protons by $1^{\circ}$ or more, and up to $20$--$30^{\circ}$. This implies that the deflection in the intergalactic medium must be taken into account in order to identify the sources of UHECRs.
\end{abstract}

\maketitle

The identification of the sources of ultra-high energy cosmic rays (UHECRs) is one of the central problems of astroparticle physics. No strong signatures of sources have been seen in the data so far -- the observed UHECRs show a surprisingly high level of isotropy, with no significant small scale clustering, see e.g.~\cite{Kuznetsov:2020hso} for a recent discussion or~\cite{Kachelriess:2019oqu} for a review~\footnote{Indeed, at  ``intermediate'' energies {$E \geq 8\cdot 10^{18}$}~eV, a dipole anisotropy at the level of 6\%  have been detected~\cite{Aab:2017tyv}; at $E>5.7\cdot 10^{19}$~eV, there is evidence (with a $3.4\sigma$ significance) for a hot spot with radius $\sim 25^{\circ}$~\cite{Abbasi:2014lda}. For the map of all 266 observed events with $E>5.7\cdot 10^{19}$~eV, see e.g.~\cite{Matthews:2017waf}.}.

This absence of small scale clustering is believed to arise from the deflection of UHECRs in magnetic fields during their propagation between the sources and Earth. This effect depends on the mass composition of UHECRs, which is not known -- the same magnetic field deflects heavy nuclei much more strongly than protons. In general, the total deflection of UHECRs can be separated into the deflection by the Galactic magnetic field (GMF) and by the intergalactic magnetic field (IGMF). For the GMF, both estimates~\cite{Kachelriess:2019oqu} and numerical studies~\cite{Giacinti:2011uj} predict average deflection angles at the level of $1^{\circ}$ for protons with energy $5\times10^{19}$~eV outside of the galactic plane. However, these predictions still suffer from many uncertainties in GMF models (see~\cite{2012ApJ...757...14J,2019Galax...7...17H,Kuznetsov:2020hso}, for a discussion). There is a hope that a reliable model of the GMF would make it possible to identify the sources of the UHECRs by re-tracing protons (see e.g.\cite{2019Galax...7...17H, Boulanger:2018zrk}), which would constrain the mass composition of the highest energies UHECRs~\citep{Kuznetsov:2020hso}.
  
In the approach outlined above, it is often assumed that the deflection of UHE protons in the IGMF is not very large (below 1 degree), and that it is sufficient to re-trace only the large-scale component of the GMF, thereby neglecting the IGMF contribution. If instead the deflection of protons by the IGMF is at the level of a few degrees or more, this would make the identification of the sources much more difficult -- a realiable model of the magnetic field strength and topology in the local Universe would be required.

\begin{figure*}
    \centering
    \includegraphics[width=\textwidth,trim={0 3.5cm 0 0},clip]{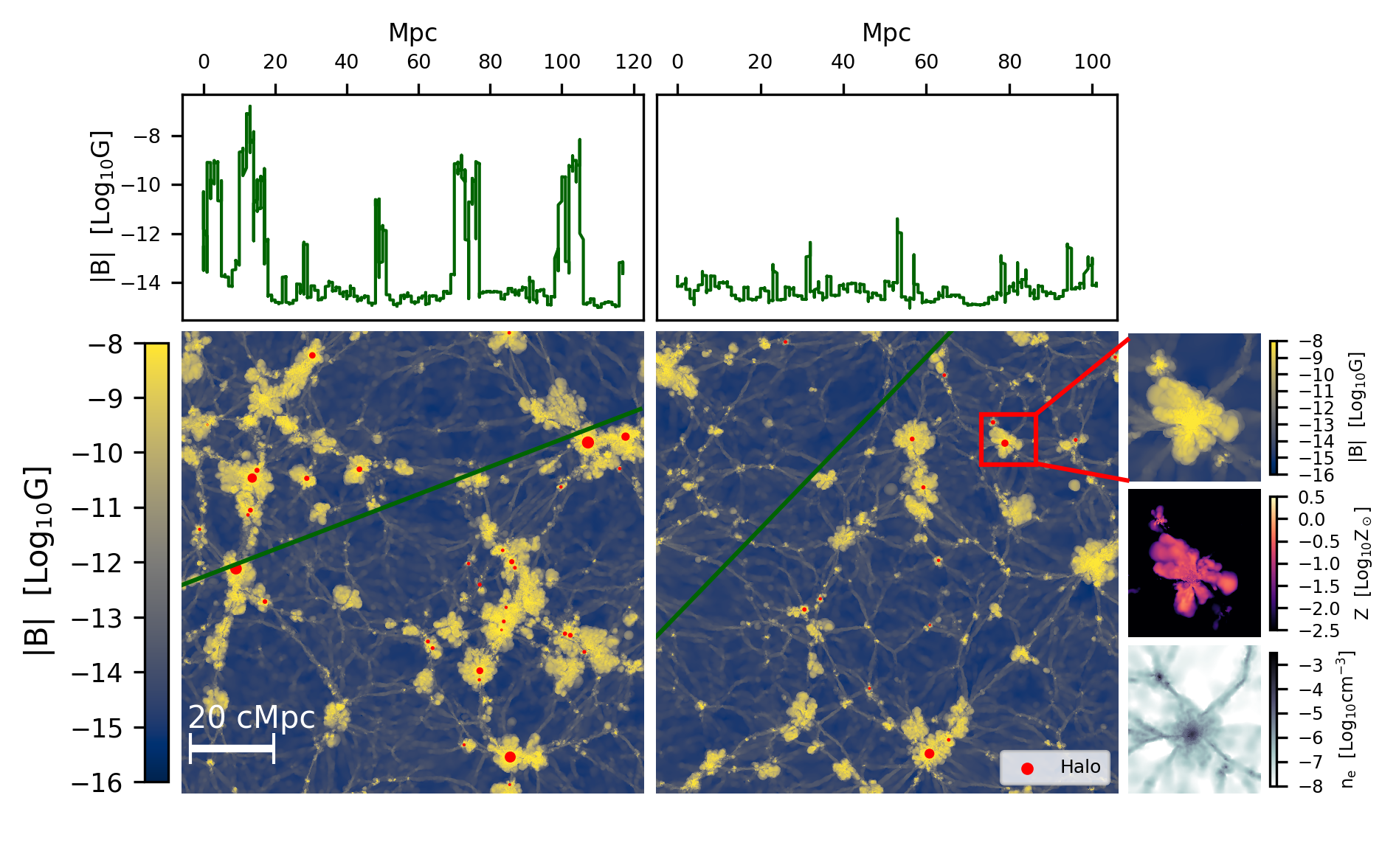}
    \caption{Maps of the magnetic fields for two random slices of the TNG100 simulation (20 kpc deep). The red circles show dark matter halos with total mass above $10^{11.5}\MSUN$ which reside within each slice, with radii corresponding to $1.5$ times their virial radius. The magnetic field values along two random lines of sight (green lines in maps) are shown in the upper panels. We also zoom on a region with an extended, magnetized bubble and show in detail magnetic field magnitude (upper right panel), gas metallicity (middle right panel), and free electron number density (bottom right panel). 
    }
    \label{fig:zoom}
\end{figure*}

Existing numerical cosmological simulations do not give consistent predictions about the strength of IGMFs \cite{Sigl:2004yk,Dolag:2004kp,Hackstein:2016pwa,Hackstein:2017pex}. For example,~\cite{Sigl:2004yk}
find that the IGMF deflection is more important than the GMF contribution, while the simulation performed in~\cite{Dolag:2004kp} that used a seed magnetic field $B\sim 2\times10^{-12}$~G produces significantly smaller deflection angles, which are negligible in comparison to the GMF contribution. More recent simulations~\cite{Hackstein:2016pwa,Hackstein:2017pex} have studied UHECR propagation using the cosmological code \textsc{Enzo}, finding that the magnetic fields in voids have a minor influence on the propagation of UHECRs. \cite{AlvesBatista:2017vob} adopt the maximum possible seed magnetic field strength based on experiment and observational inferences, $B\sim 10^{-9}$~G, and derive deflections of the order of ten degrees. Overall, the dependence of these results on the unknown initial conditions for cosmic magnetic fields prevents us from deriving robust conclusions.

In this paper, we concentrate on the part of the IGMFs that may be, to a large extent, independent of the initial magnetic field seeds. We have recently pointed out in~\cite{Garcia:2020kxm} that, according to the IllustrisTNG cosmological simulations~\cite{nelson18,springel18,pillepich18,naiman18,marinacci18}, baryonic feedback processes, such as AGN and supernovae-driven outflows, create extended over-magnetized bubbles in the intergalactic medium (IGM). These have magnetic field strengths that can be many orders of magnitude larger than those in similarly dense regions of the IGM, and are determined by processes that take place inside galaxies rather than by the initial properties of the magnetic fields. The contribution to the deflection of UHECRs from these bubbles can be considered as a minimal effect of the IGMFs, present even if the seed fields are feeble. Importantly, the IllustrisTNG simulations have been shown to return overall properties of the galaxy populations in good agreement with data across a wide range of observables and regimes. In comparison to previous numerical experiments, this provides support for the emerging properties of AGN and stellar feedback, that in turn are responsible for the over-magnetized bubbles that are inflated well into the IGM.

In this letter, we show that, according to the IllustrisTNG model, the deflections of the highest-energy protons by the IGMF can be split into two comparable groups -- below one degree and 1--30 degrees. Such a picture suggests that while the fraction of UHECRs with relatively small deflections in IGMF is significant enough to leave the possibility of source identification, the effect of IGMFs is strong and should be taken into account. 

\textit{Simulations}.---The IllustrisTNG project is a set of cosmological gravo-magnetohydrodynamic simulations (hereafter TNG,~\cite{nelson18,springel18,pillepich18,naiman18,marinacci18}) based on the moving-mesh \textsc{Arepo} code \cite{springel2010MNRAS.401..791S} that adopt Planck 2015 cosmological parameters~\cite{Plank2016A&A...594A..13P}. This code solves the coupled equations of self-gravity and ideal MHD~\cite{2011MNRAS.418.1392P,2013MNRAS.432..176P}. Specifically, in this work, we use the publicly available TNG100-1 simulation (hereafter TNG100~\cite{Nelson2019ComAC...6....2N}), that is the highest resolution run of the TNG100 series, with a homogeneous initial seed magnetic field with strength $10^{-14}$~cG (comoving Gauss). TNG100 has a box size $\sim (110~\text{cMpc})^3$ and contains $1820^3$ dark matter particles and an equal number of initial gas cells with mass resolution of $m_{\text{DM}} = 7.5 \times 10^6~\MSUN$ and $m_{\text{bar}} = 1.4 \times 10^6~\MSUN$, respectively.

\begin{figure}
    \centering
    \includegraphics[width=\linewidth]{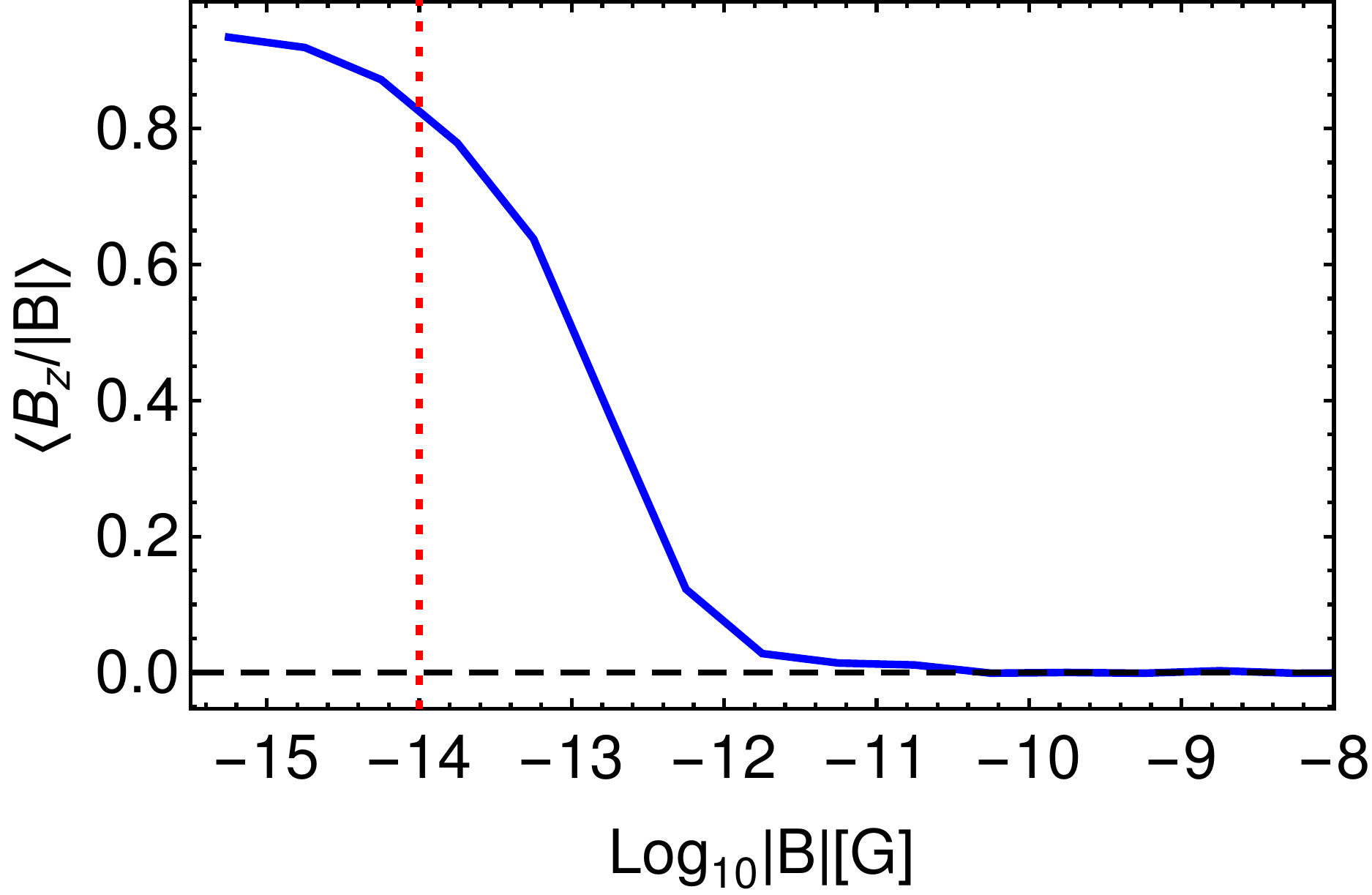}
    \caption{Loss of the ``memory'' of the initial magnetic seed field in TNG100. The blue curve shows the average orientation of the $z=0$ magnetic field along the $z$ direction of the simulation domain (i.e. the orientation of the initial uniform seed) as a function of the $z=0$ field strength, for regions with $n_e < 200\langle n_e \rangle$. Similar results hold for other electron number densities. The red vertical line marks the seed field strength.
    }
    \label{fig:Borientation}
\end{figure}

\begin{figure}
    \centering
    \includegraphics[width=\linewidth]{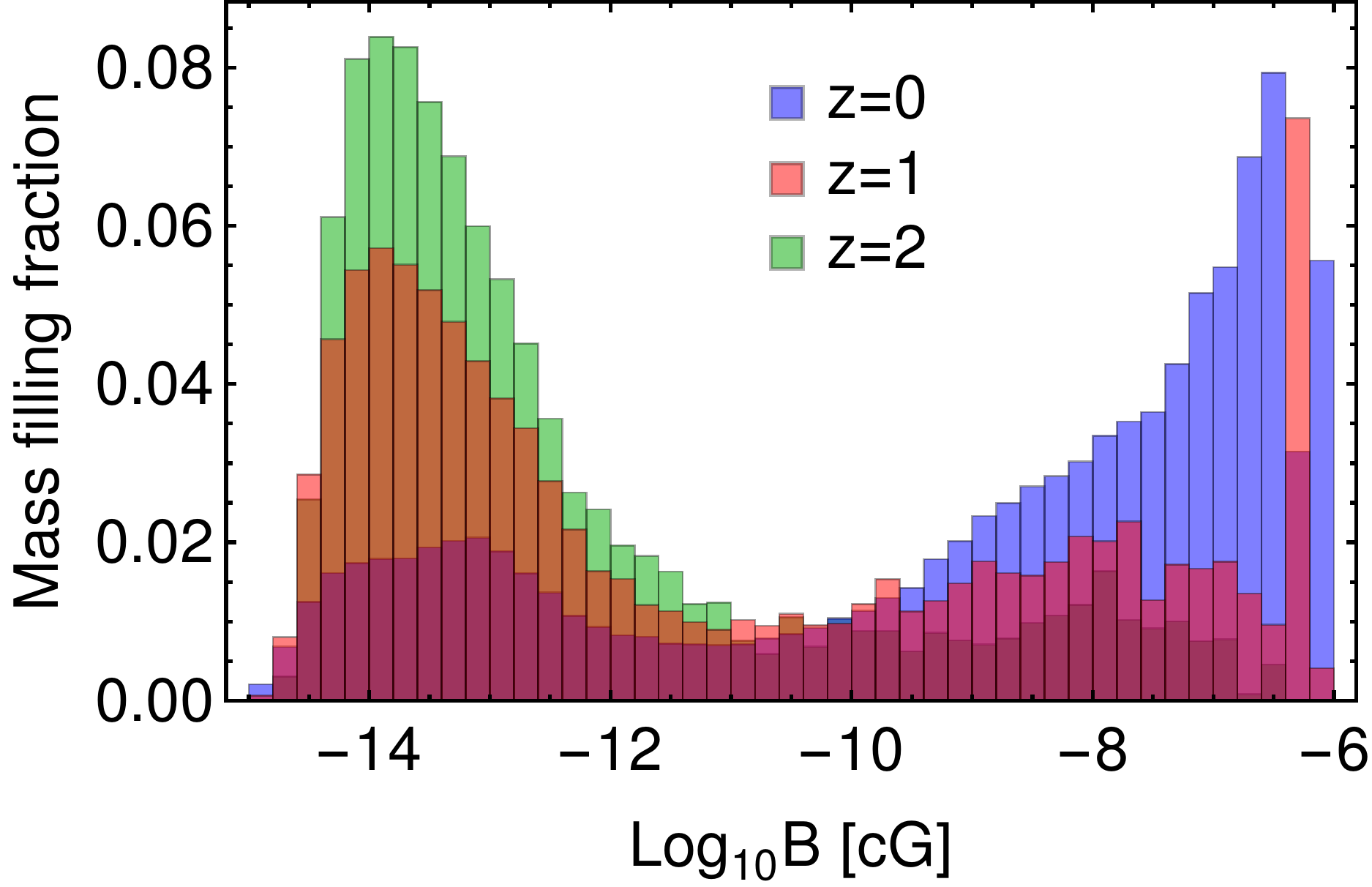}
    \caption{Amplification of the magnetic fields in TNG100 with time. We show the mass weighed distribution of comoving magnetic field strength at redshifts $z=0$, $1$, and $2$.}
    \label{fig:Bevolution}
\end{figure}

The TNG simulations adopt a detailed galaxy formation model that is important for the resulting structure of
the magnetic fields both inside and outside galaxies and clusters. For a more detailed description of this model, see Section 2 in~\cite{Garcia:2020kxm} and references therein. Of particular importance for the magnetic field evolution is the inclusion of AGN feedback and galactic winds launched by supernovae, among other astrophysical processes~\citep{Weinberger2017MNRAS.465.3291W,Pillepich2018MNRAS.473.4077P}. Apart from adiabatic contraction, magnetic fields in TNG experience strong amplification by small-scale dynamos, which are then distributed beyond galaxies and even halos by energetic outflows.

\textit{Outflow-driven bubbles}.---As shown in~\cite{Garcia:2020kxm} the TNG simulations predict that for $z \lesssim 2$ as galaxies form, and AGN and stellar feedback become important, these processes form large regions of strong magnetic fields far outside of the virial radii of collapsed structures, thus creating over-magnetized bubbles. These large-scale bubbles contain magnetic fields that are several orders of magnitude stronger than in the unaffected regions of the IGM with the same electron density. In particular, if we identify bubbles as regions with $B>10^{-12}$~cG, enhanced metallicity, and with clear outflowing kinematic signatures, they can be as large as tens of Mpc -- see Fig.~\ref{fig:zoom}.

The properties of these magnetic bubbles are determined by baryonic feedback physics. It was demonstrated in~\cite{Garcia:2020kxm} that, similarly to the magnetic fields inside galaxies, magnetic fields in the bubbles are to a certain extent independent of the initial conditions for magnetic fields assumed in the simulation. In Fig.~\ref{fig:Borientation} we show how the simulated magnetic field ``forgets'' about the initial orientation of the seed magnetic field, which is along the $z$ axis, for increasing field strengths, i.e. due to amplification. It is clearly seen that for outflow-driven bubbles (e.g. for $B>10^{-12}$~cG), there is no longer a preferred direction of the field.

The TNG100 simulation suggests that such over-magnetized bubbles form quite recently in cosmic history, at redshifts $z \lesssim 2$, see Fig.~\ref{fig:Bevolution}. While at $z=2$ most of the gas mass still has initial-like magnetic field strengths ($10^{-14}$~cG for TNG100), at lower redshifts the situation changes. 
At $z=0$ the bubbles are not rare, occupying more than $10\%$ of cosmological volume for magnetic fields stronger than $10^{-12}$~G and more than $3\%$ for $B>10^{-9}$~G~\cite{Garcia:2020kxm}. 

\begin{figure*}
    \centering
    \includegraphics[width=\textwidth,trim={0 0.2cm 2.5cm 1.5cm},clip]{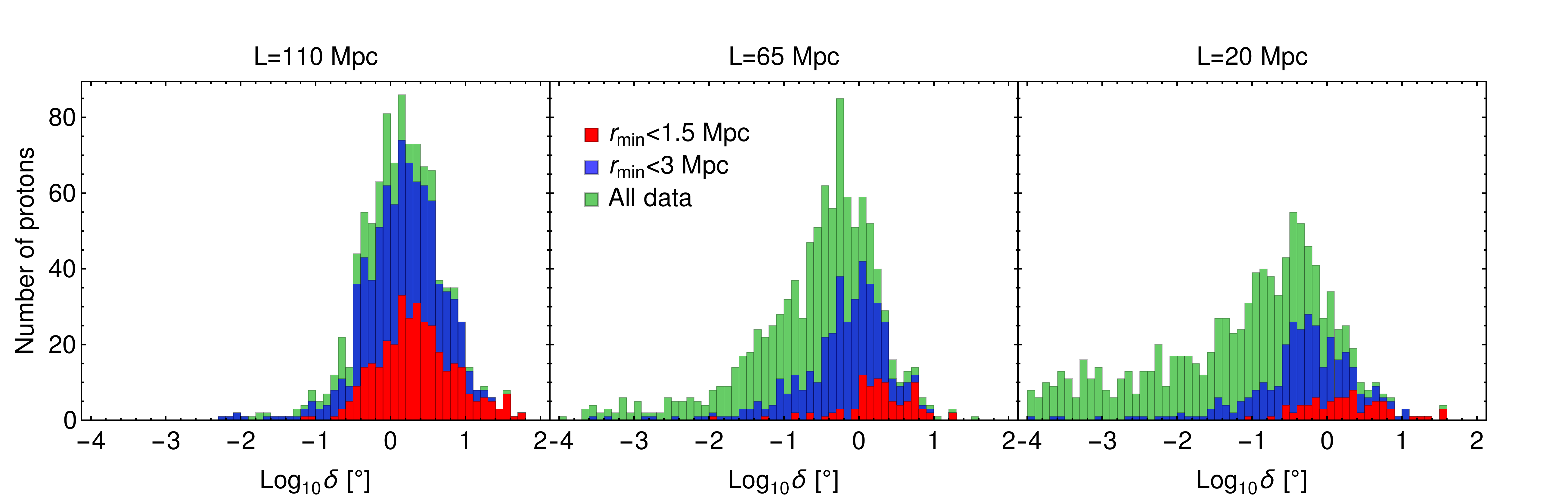}
    \caption{Distribution of the deflection angles $\delta$ for 1000 protons with energy $10^{20}$~eV with random initial position in our fiducial volume in the TNG100 simulation at $z=0$ for trajectory lengths of 110, 65, and 20 Mpc, from left to right. In each panel, colors shows cuts by $r_{\min}$, the distance to the closest halo with mass above $10^{12.5}M_{\odot}$ along the trajectory.}
    \label{fig:deflection_IGMF}
\end{figure*}

\begin{figure}
    \centering
    \includegraphics[width=\linewidth,trim={0 0.3cm 0 0},clip]{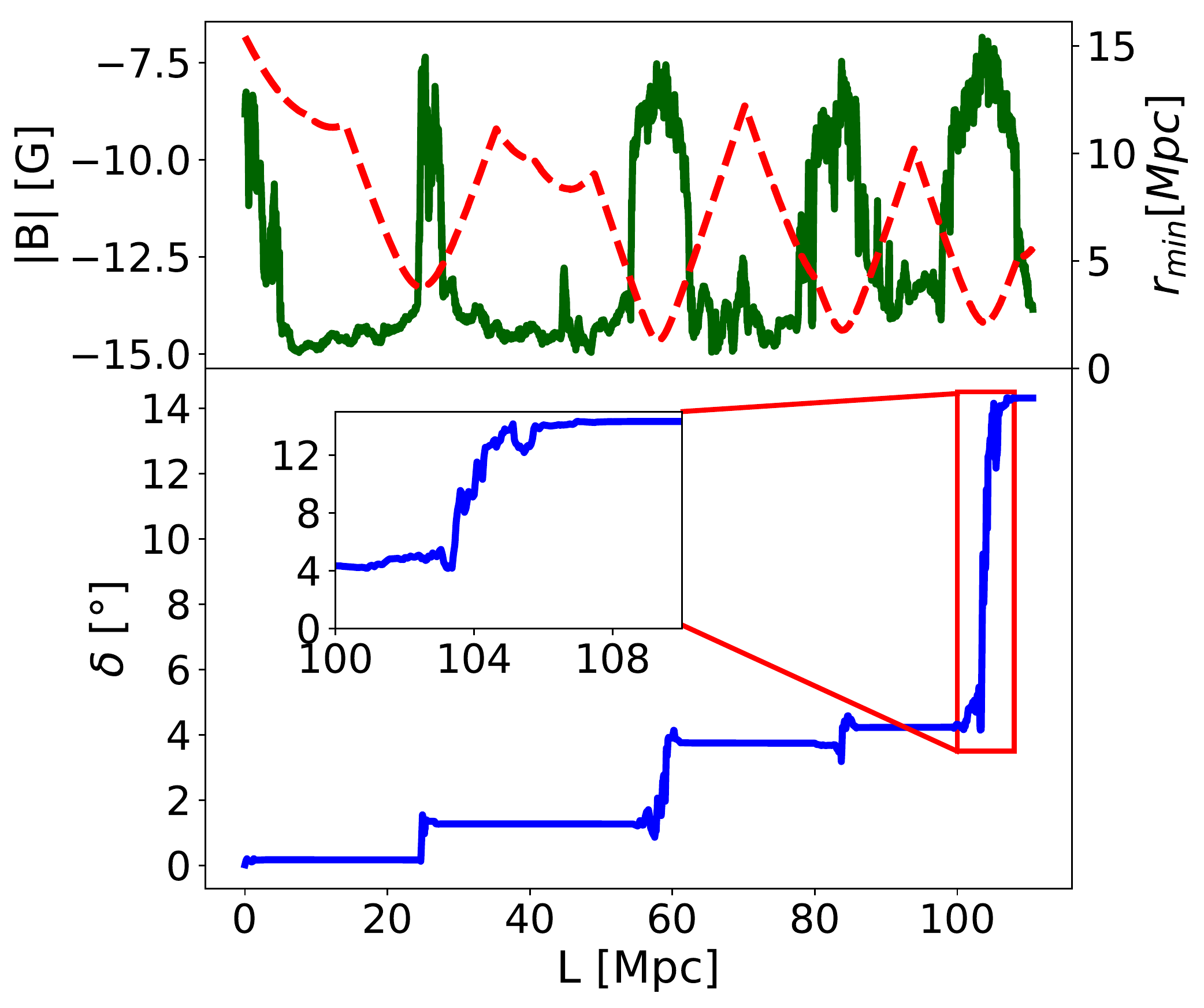}
    \caption{Evolution of the magnetic field strength $|B|$ (top panel, green curve), the minimal distance $\mathrm{r_{min}}$ to a massive halo $>10^{12.5} \mathrm{M_\odot}$ (top panel, red dashed curve), and of the corresponding deflection angle $\delta$ (bottom panel), along a random trajectory for a proton with energy $10^{20}$~eV.}
    \label{fig:deltachange}
\end{figure}

\textit{Deflection of UHECRs}.---To study the effect of the over-magnetized bubbles on the propagation of UHECRs we trace trajectories of high-energy protons in the TNG100 simulation. The change of the velocity of a particle traversing a magnetic field is given by
\begin{equation}
    \Delta \bm{v} = \frac{q}{E_p} \int [\bm{v}\times \bm{B}] dl,
    \label{eq:Deltap}
\end{equation}
where $q$ and $E_p$ are the charge and energy of the proton and the integral is taken along the trajectory of a particle.

We take the magnetic field configuration from a random $(30\text{ Mpc})^2\times 110$~Mpc sub-volume of TNG100, with the long side parallel to the direction of the initial seed field, averaging within $(40\text{ kpc})^3$ voxels. We then track the propagation of protons along the long side of this volume, following their changing velocity and direction according to Eq.~\eqref{eq:Deltap}. We consider 1000 protons with energy $10^{20}$~eV, with initial positions randomly chosen within a $20\text{ Mpc}\times 20\text{ Mpc}$ area at the volume boundary. Initial velocities are oriented along the long side of the volume, and we propagate each until the end of the volume.

The resulting distribution of the angle between initial and final propagation direction is shown in Fig.~\ref{fig:deflection_IGMF}, for integration along paths of 110, 65, and 20 Mpc (left to right). For path lengths of $\sim$\,110 Mpc, about $2.8\%$ of these protons are deflected by $\delta< 0.1^{\circ}$. However, another $35.5\%$ of protons have \mbox{$0.1^{\circ} <\delta < 1^{\circ}$}: these deflections are much larger than the $\delta \sim 0.001^{\circ}$ that would be expected from a seed field of $B\sim 10^{-14}$~cG along the whole 110~Mpc trajectory. Finally, $61.7\%$ of protons encounter even larger magnetic fields both in bubbles and sometimes inside halos and attain $\delta > 1^{\circ}$, where deflections can reach up to tens of degrees. In fact, the deflection angles are larger for protons which pass close to massive halos, particularly for shorter path lengths, as shown in the different histogram colors.

An example of the deflection angle evolution of a test protons is shown in Fig.~\ref{fig:deltachange}. We see that most of the deflection is acquired during the $\sim 2.5$~Mpc-size region that corresponds to the crossing of a magnetic bubble.

\textit{Conclusions and discussion}.---We find that, according to the IllustrisTNG model of galaxy formation, the influence of intergalactic magnetic fields  on the propagation of the UHECRs is essential and must to be taken into account when searching for the sources of these particles. The average value of the magnetic field strength in voids strongly depends on the (unknown) initial seed magnetic fields. However, large, outflow-inflated, over-magnetized bubbles form around collapsed halos, as recently emphasized in~\cite{Garcia:2020kxm}. Magnetic fields in these regions, which can extend out to tens of Mpc, are created by feedback processes in the innermost regions of galaxies and can be orders of magnitude larger than in unaffected regions of the IGM with the same baryon density. The strength and geometry of these fields, being to great extent defined by processes occurring within galaxies, ``forget'' the initial conditions  of the seed field. As a result, they provide a non-negligible contribution of the IGMFs to UHECR deflection, even for extremely small values of the initial seed.

Our results can potentially explain why the sources of UHECRs have not been identified yet and, at the same time, suggest a possible promising way forward to resolve this problem. Our study finds that about half of all protons with $10^{20}$~eV energy are deflected in the IGM by $\delta < 1^{\circ}$ along a $\sim100$ Mpc path. At the same time, there is a significant fraction of protons that go through over-magnetized bubbles and are strongly deflected. This means, for example, that among strongly deflected particles, identified by the procedure of~\cite{Kuznetsov:2020hso} as heavy nuclei, there may also be a fraction of protons that passed through, or nearby, such regions.

In order to explicitly account for these bubbles and so help identify the sources of UHECRs, we plan to perform a constrained simulation of the local volume, extending to at least $100$-$200$~Mpc. This simulation will reproduce the observed large-scale structure around the Milky Way, predict the corresponding outflow-driven bubbles, and create a large-scale map of the local magnetic fields.

Such a volume would contain most of the possible sources of the highest energy CRs that could reach us despite the GZK cut-off. It would be, of course, difficult to use a simulated map for re-tracing the UHECRs, as e.g. the details of the geometry of magnetic fields inside the bubbles depend on the kinematics of outflows. Even without explicit re-tracing, it would assist statistical inferences on the fraction of protons which remember their original directions when entering the Galaxy. Such a constrained simulation will also enable detailed mock predictions for the detection of the local bubbles, through for instance Faraday Rotation Measure (RM). As we expect the RM in the bubbles to be much larger than in voids, but still lower than in the collapsed structures or even in filaments (see~\cite{Garcia:2020kxm} for a discussion), such a detection would require a non-trivial statistical analysis.

We thank Dmitry Semikoz, Volker Springel, J.-P. Kneib and T.Theuns for very fruitful discussions.
AS are supported by the FWF Research Group grant FG1. AA, KB and AB are supported by the European Research Council (ERC) Advanced Grant ``NuBSM'' (694896).
 
\bibliography{refs.bib}

\end{document}